\documentclass[conference]{IEEEtran}
\usepackage{titlesec}
\ifx\subparagraph\undefined
  \titleclass{\subparagraph}{straight}[
    section]
\fi
\IEEEoverridecommandlockouts
\usepackage{cite}
\usepackage{hyperref}
\usepackage{amsmath,amssymb,amsfonts}
\usepackage{algorithmic}
\usepackage{graphicx}
\usepackage{eurosym}
\usepackage{textcomp}
\usepackage{xcolor}
\usepackage{xargs}
\usepackage{color,listings}
\usepackage{tabularx}
\usepackage{tabularray}
\usepackage{listings}
\usepackage{caption}
\usepackage{multirow}
\usepackage{ragged2e}
\usepackage{url}
\usepackage{subcaption}
\usepackage[colorinlistoftodos,prependcaption,textsize=scriptsize]{todonotes}
\def\BibTeX{{\rm B\kern-.05em{\sc i\kern-.025em b}\kern-.08em
    T\kern-.1667em\lower.7ex\hbox{E}\kern-.125emX}}

\newcommandx{\mitra}[2][1=]{\todo[linecolor=pink,backgroundcolor=pink!25,bordercolor=pink,#1]{\textbf{Mitra comments: }#2}}
\newcommandx{\pragyan}[2][1=]{\todo[linecolor=green,backgroundcolor=green!25,bordercolor=green,#1]{\textbf{Pragyan comments: }#2}}

\newcommandx{\citeme}[2][1=]{\todo[linecolor=red,backgroundcolor=red!25,bordercolor=red,#1]{\textbf{CITE} #2}}
\newcommandx{\fillin}[2][1=]{\todo[linecolor=red,backgroundcolor=red!25,bordercolor=red,#1]{\textbf{FILL IN} #2}}

\begin{document}

\title{From Community Forums to Issue Trackers: A Moodle Case Study}

\author{\IEEEauthorblockN{Pragyan K C\textsuperscript{1}, Mitra Bokaei Hosseini\textsuperscript{1}}
\IEEEauthorblockA{\textsuperscript{1}University of Texas at San Antonio, San Antonio, TX, USA\\
\textit{
[pragyan.kc, mitra.bokaeihosseini]@utsa.edu}
}
}

\maketitle

\begin{abstract}

Sustaining open-source software (OSS) requires effective practices for evolution and change management. In OSS projects, evolution is largely driven by feature requests and enhancements proposed by diverse stakeholders. These requests are often discussed across multiple communication channels, particularly community forums and issue trackers, where stakeholders negotiate intent, clarify requirements, and coordinate development.  
Despite prior research on OSS forums and issue trackers, we lack an empirical understanding of who creates and maintains links between forum posts and tracker issues, and how these links support clarification, feedback, and coordination throughout feature request lifecycles. 
To address these questions, we conduct an in-depth case study of Moodle, a widely used open-source learning management system. Our study combines (1) an empirical analysis of cross-channel trace links between Moodle's community forum and its Jira issue tracker, (2) semi-structured interviews with developers, 
and (3) semi-structured interviews with forum participants. 
Our results show that cross-channel traceability is rare: only 818 of 23{,}169 ($\approx$3.5\%) feature request issues in Moodle’s Jira link back to a community forum, and authorship differs by channel, with developers authoring 52.8\% of tracker issues, while forum feature requests are predominantly authored by users, and only 230 linked pairs share the same author. The qualitative findings further reveal that the transition from forum posts to issues is largely ad hoc, with limited tool support and unclear role ownership, and that users often experience the process as opaque or weakly responsive. 

\end{abstract}



\section{Introduction}


Many organizations have embraced open source software (OSS), realizing substantial benefits in cost savings, accelerated development, and interoperability through open standards~\cite{nagy2010organizational,chesbrough2023measuring,osisurvey}. According to the 2025 State of Open Source Report, 96\% of organizations increased or maintained their use of OSS in the past year, with over a quarter (25.71\%) reporting a significant increase~\cite{osisurvey}. 
This rapid growth has also exposed persistent challenges in the maintenance and evolution practices that can significantly drive up costs and erode user satisfaction, ultimately threatening the long-term success and adoption of OSS~\cite{osisurvey,coelho2017modern,avelino2019abandonment,coelho2020github,costal2015aligning,schueller2024modeling,siena2014modelling,malcher2024investigating}. Evolution is largely driven by feature requests proposed by diverse stakeholders~\cite{pragyan2025demystifying}. Successfully managing these requests requires frequent, direct communication—often involving clarification and negotiation—among requesters, developers, maintainers, and project owners~\cite{hellman2022characterizing}. Effective communication yields a shared understanding of feature requests and the requester’s underlying intent among stakeholders~\cite{hellman2022characterizing}. 
One such communication channel for requesting features in OSS is the community forums, where the stakeholders of OSS may come to engage with the larger
community~\cite{hellman2022characterizing}. 
Besides community forums, OSS projects also rely on tool-supported issue trackers (e.g., GitHub or Jira) as another channel to report, track, and manage feature requests~\cite{bissyande2013got,raatikainen2022improved}. 

Prior work has examined OSS community forums and issue trackers largely as separate channels. Forum studies characterize participation and communication behaviors and role differences in OSS forums, and explore mining online discussions for requirements and user needs~\cite{hellman2022characterizing,nugroho2021project,kanchev2017canary}. In parallel, issue tracker research emphasizes trackers as central artifacts for coordinating work and evolution, addressing challenges such as collaboration and information overload, modeling and managing relationships among issues (e.g., dependencies), and recovering dependency links via information-retrieval techniques~\cite{bertram2010communication,baysal2014no,heck2013analysis,borg2014recovering,raatikainen2022improved}. More recently, Tizard et al.~\cite{tizard2022software} analyzed the HTML links embedded in community forum threads for two OSS projects to characterize how forum discussions reference issue trackers. 

Despite the prior research, it remains unclear who creates and maintains traceability links between forum posts and tracker issues, and—critically—how these trace links are used in practice to support feedback and clarification loops for feature requests. In particular, we lack evidence on whether original forum requesters are engaged for follow-up (e.g., clarification questions, negotiations, validation, status updates), whether they continue participation on the linked tracker issue, and whether developers consult linked forum posts to recover context and rationale or instead rely primarily on the tracker artifact and their own intuition.  

To address these gaps, this paper aims to empirically investigate how traceability links between community forums and issue trackers are established, maintained, and used in OSS. We operationalize this investigation through an in-depth case study of Moodle, an open-source learning management system (LMS)\cite{moodleDotOrg}, as the definitive example of large-scale OSS, due to its massive codebase, its global infrastructure of contributors, and its dominant market share in the education sector~\cite{edzlms2024moodle}.
This paper makes three contributions:
\begin{itemize}
\item \textbf{An empirical analysis of cross-channel traceability between community forums and the Jira issue tracker in Moodle.} We provide evidence of a substantial traceability gap, showing that only 818 of 23,169 feature request issues ($\approx$3.5\%) explicitly link back to community forum discussions.
\item \textbf{An empirical characterization of stakeholder participation in feature request evolution across communication channels.} Our findings reveal distinct authorship patterns, with forum feature requests being predominantly user-authored, while issue tracker feature requests are more frequently authored by developers, highlighting different stakeholder roles in requirements communication and evolution.
\item \textbf{An investigation of how feature requests are transformed as they move from community discussions to development artifacts.} We show that issue descriptions are generally more technical and less readable than their originating forum discussions, while interviews with developers and users reveal that the transition between channels is largely ad hoc, supported by limited traceability mechanisms and unclear ownership of request handoffs.
\end{itemize}

The remainder of this paper is structured as follows. Section~\ref{sec:relatedwork} reviews the background \& related work. Section~\ref{sec:researchMethod} describes the case study design. Section~\ref{sec:results} presents the results, followed by discussion in Section~\ref{sec:disscussion}. Section~\ref{sec:threats} outlines threats to validity, and Section~\ref{sec:conclusion} concludes the paper.

\section{Background \& Related Work} \label{sec:relatedwork}


\subsection{Communication Channels in OSS}

\noindent\textbf{Community Forums:} Researchers have identified requirements-relevant information in user feedback in several online channels, including app stores, social media, and product forums~\cite{tizard2019can,pagano2013user,guzman2017exploratory,malcher2024investigating,tizard2022voice}. 
When a software user posts in a product forum, they often describe an issue they have encountered and are seeking help to resolve~\cite{tizard2020voice}. 
Experienced users, as well as members of the development team, can reply to forum posts giving guidance on the nature of the user’s issue~\cite{tizard2019can}. Daniel et al. further argued that some of the most impactful forms of external engagement include task-related requests and forum discussions, both of which represent essential channels for influencing project evolution~\cite{daniel2020impact}.
Previous studies also highlight the need for automated analysis tools to assist in requirement extraction from such communication channels, as manual extraction can be extremely time-consuming due to the large volume of feedback~\cite{guzman2016needle,tizard2019can,guzman2017little,maalej2015bug,di2017surf,panichella2016ardoc,chen2014ar,khan2019analysis,maalej2025automated,gao2018online}. 
Harman et al. showed the importance of addressing user
concerns, finding a strong correlation between customer ratings and the popularity of mobile applications~\cite{harman2012app}. Tizard et al. showed that product forums differ from app reviews in both content and structure and may require alternative analysis techniques~\cite{tizard2019can}. Recent work has further shown that analyzing forum feedback at finer granularities (e.g., individual sentences) can miss important context that is only visible when examining complete posts and their evolution within a thread~\cite{wang2024conversation}.

\noindent\textbf{Issue Trackers:} Issue trackers play a central role in OSS development, functioning not only as repositories for managing work items but also as important socio-technical spaces that facilitate interaction and coordination among contributors~\cite{chrupala2012learning,zhou2011does,montgomery2025mining}. Their usage patterns and dynamics have been shown to meaningfully influence project productivity and overall development outcomes~\cite{montgomery2025issue,ruvimova2022exploratory}. Maalej and Happel observed that issue trackers frequently serve as the primary arena for external contributor participation~\cite{maalej2010can}. 

Prior research has examined how issue tracker activities reflect project health and development processes~\cite{crowston2003defining,gousios2008measuring,robles2014estimating,zhao2024openrank,shen2025external}. 
While these studies highlight the significance of external contributions to OSS projects, many focus primarily on aggregate indicators—such as the number of contributors~\cite{setia2012peripheral} or commit counts~\cite{lee2020role}—without deeply examining the nature and substance of contributor activities within issue trackers. Further, these works analyze the communication channels in isolation—either mining user feedback or quantifying developer activity. In contrast, our study investigates how feature requests move across these communication channels. 



\subsection{Traceability in OSS Communication Channels} 
Prior research has examined various types of trace links within issue trackers~\cite{montgomery2025issue,luders2023understanding}, with duplicate detection emerging as a central challenge~\cite{anvik2005coping,deshmukh2017towards,he2020duplicate,wang2008approach,borg2013analyzing,li2018issue}. 
Tomova et al.~\cite{tomova2018use} further showed that the rationale for selecting specific link types is not always clear across OSS projects.

Prior research has explored automatically linking user feedback to issue tracker artifacts at scale~\cite{pilone2025automatically}. For instance, Haering et al.~\cite{haering2021automatically} showed that app store reviews containing bug reports can be automatically matched to corresponding issue tracker entries. Tizard et al.~\cite{tizard2022software} examined the product forums of two large OSS projects (VLC and Firefox) and found that users frequently report previously undocumented requirements that are later transferred to issue trackers by \textit{contributors.} Their quantitative analysis revealed that forum posts are often manually linked to issue tracker entries and documentation, and they further applied deep-learning models to automatically identify such cross-channel links. Through their manual analysis, for both VLC and Firefox, they identified that a \textit{significant number of links} were posted in forums directing to the issue trackers and product documentation.

While previous work investigates link patterns across different OSS communication channels and explores automated matching techniques, it primarily focuses on detecting and characterizing cross-channel references. In contrast, our study examines how feature requests originate, evolve, and transform across communication channels, analyzing stakeholder roles, traceability prevalence, linguistic reformulation, prioritization practices, and user experiences. Rather than focusing solely on link detection, we expose a broader socio-technical traceability gap and show that the transition from user-facing discussions to development artifacts is largely ad hoc and inconsistently supported in OSS ecosystems.

\begin{figure*}[t]
  \centering
    \fbox{
  \begin{subfigure}[t]{0.50\textwidth}
    \centering
    \includegraphics[width=\linewidth]{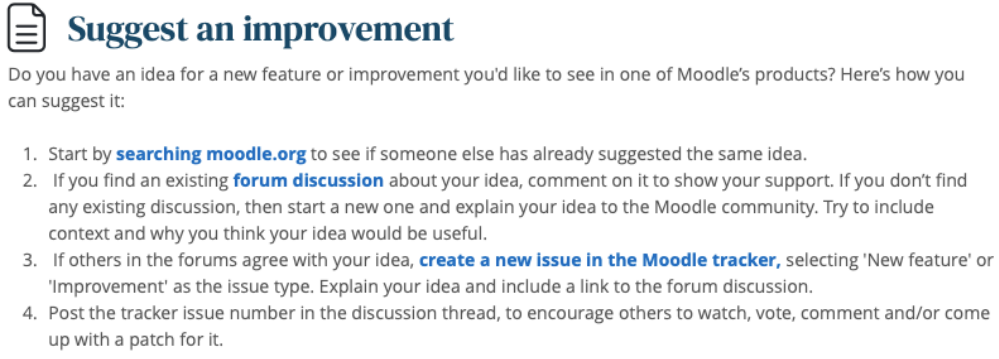}
    \caption{Instructions for submitting a feature request}
    \label{fig:howtosubmitarequest}
  \end{subfigure}
  \hfill
  \begin{subfigure}[t]{0.40\textwidth}
    \centering

    \begin{subfigure}[t]{0.49\linewidth}
      \centering
      \includegraphics[width=\linewidth]{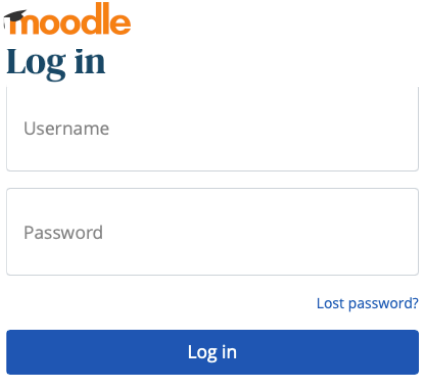}
      \caption{Login for Moodle}
      \label{fig:moodlelogin}
    \end{subfigure}
    \hfill
    \begin{subfigure}[t]{0.49\linewidth}
      \centering
      \includegraphics[width=\linewidth]{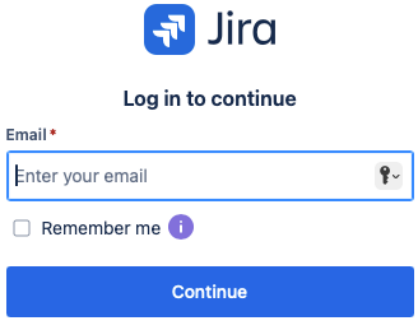}
      \caption{Login for JIRA}
      \label{fig:JIRAlogin}
    \end{subfigure}

  \end{subfigure}}

  \caption{Feature request submission instructions and multiple channels in Moodle}
  \label{fig:howToSubmitRequest}
\end{figure*}

\section{Case Study Design} \label{sec:researchMethod}
For our case study, we choose Moodle, an open-source LMS, that has over 507 million registered users worldwide and more than 54 million active courses~\cite{moodleStats2026}. Further, it is active in 235 countries (essentially every country with internet access), indicating its global reach~\cite{moodleStats2026}. 
Moodle consists of millions of lines of code. It is a modular monolith, meaning that while it is one large system, it relies on over 2,000 community-contributed plugins to function at scale~\cite{moodle2024opensource}. Over 1,000 developers contribute to the core code, coordinated by ``Moodle HQ.'' This is significantly larger than most open-source projects, which often rely on fewer than 10 core maintainers~\cite{moodle2026about}. It is translated into over 100 languages, requiring a massive localization infrastructure~\cite{moodle2026about}. 

\begin{figure*}[]
  \centering
  \fbox{\includegraphics[width=0.8\textwidth]{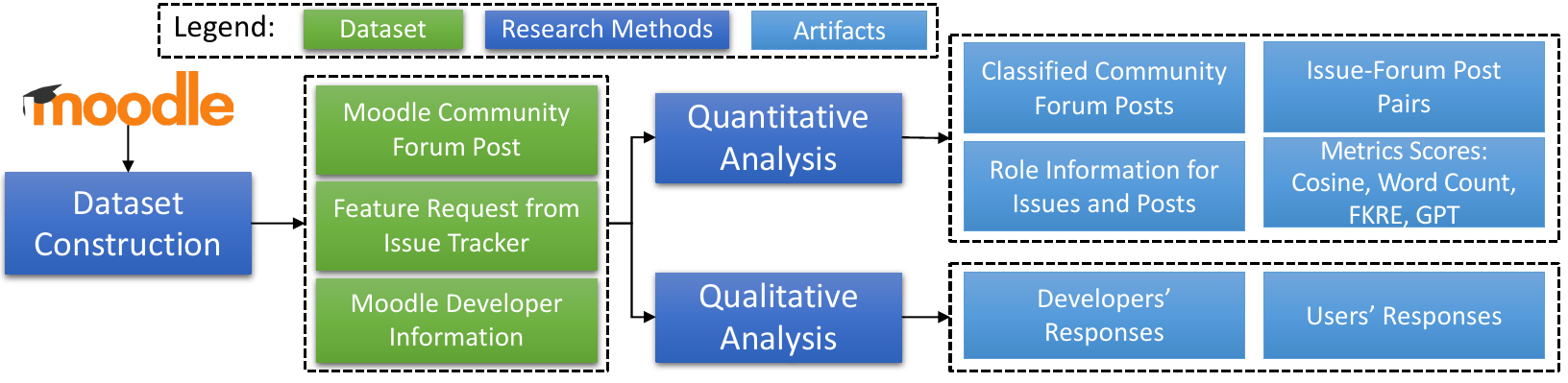}}
  \caption{Overview of the Mixed-Methods Case Study}
  \label{fig:overviewFigure}
\end{figure*}

Moodle provides multiple communication channels, including community forums and a centralized Jira issue tracker. Figure~\ref{fig:howtosubmitarequest} illustrates Moodle's recommended procedure for submitting a feature request. Users are first encouraged to search existing posts and, if needed, to start a post in the community forum to describe and refine their idea. Only after this step are users directed to submit a request in the issue tracker. Importantly, Figures~\ref{fig:moodlelogin} and~\ref{fig:JIRAlogin} show that the forum and the issue tracker are separate systems with distinct entry points and login procedures, reinforcing the separation between the two channels. 
This clear separation makes Moodle a suitable context for examining how feature requests emerge, are interpreted, and are evolved within an open-source development process. Additionally, Moodle's long development history and sustained community engagement enable observation of feature request lifecycles across multiple stages. Therefore, we investigate the traceability between \textit{issue-forum post pairs} in Moodle and the roles of stakeholders, such as users and developers, in shaping feature requests. Figure~\ref{fig:overviewFigure} summarizes our case study design~\cite{clark2008mixed}. The design begins with the construction of three datasets, which are then used in two parallel analysis tracks: quantitative and qualitative. We investigate the following research questions:


\begin{itemize}
    \item \textbf{RQ.1}: How prevalent are feature requests in Moodle's community forums?
    \item\textbf{RQ.2}: How are feature requests initiated and evolved across different communication channels, and how do different stakeholder groups contribute to the evolution and specification of requirements within these channels? 
    \item \textbf{RQ.3}: To what extent are feature-related issues in the Jira issue tracker traceable to the posts in the community forum?
    \item \textbf{RQ.4}: How do forum posts and issue descriptions differ in length, readability, and technical complexity?
    \item \textbf{RQ.5}: How do developers receive, interpret, and prioritize feature requests originating from users and community discussions?
    \item \textbf{RQ.6}: How do users submit feature requests, and how do they experience acknowledgment and feedback from developers?
\end{itemize}

\subsection{Dataset Construction} 
We construct \textbf{three} datasets that enable quantitative \& qualitative analysis and are publicly available~\cite{re2026_github_repo}.

\subsubsection{Community Forum Feature Requests Dataset}\label{dataset:CommunityForum}

We scrape data from all 62 available community forums in Moodle, yielding 204,289 posts using Beautiful Soup~\cite{BeautifulSoup}. 
The collected metadata includes the post identifier, title, publication date, author name, author email (when available), forum name, and associated comments. 
Table \ref{tab:communityForumDatasetNumbers} summarizes forum activity, presenting the five forums with the highest number of posts and the five with the lowest. The remaining forums and detailed statistics are available online~\cite{re2026_github_repo}. 

This dataset is highly imbalanced across forums, and posts span multiple purposes (e.g., help requests, troubleshooting discussions, announcements), not solely feature requests. 
To reduce bias introduced by large forums, we filter the data to retain only posts without images. Further, image content cannot be reliably processed by the text-only classification pipeline and may introduce uncontrolled variance. We then apply capped stratified random sampling\cite{blair2014applied}, treating each forum as a stratum. For each forum, up to 500 posts are randomly selected; forums with fewer than 500 eligible posts contribute all available posts. We choose a cap of 500 posts per forum as a practical trade-off between capturing within-forum diversity and maintaining computational feasibility. 
We sample 24,070 posts using this approach (see Table \ref{tab:communityForumDatasetNumbers}). 
\begin{table}[]
\centering
\caption{Statistics for \textbf{community forum dataset}}
\begin{tabular}{|lll|}
\hline
\multicolumn{1}{|l|}{\multirow{2}{*}{\textbf{Forum Name}}} & \multicolumn{2}{l|}{\textbf{\# of Posts}} \\ \cline{2-3} 
\multicolumn{1}{|l|}{} & \multicolumn{1}{l|}{\textbf{Total}} & \textbf{Sample} \\ \hline
\multicolumn{1}{|l|}{\textbf{generalHelp}} & \multicolumn{1}{l|}{70446} & 500 \\ \hline
\multicolumn{1}{|l|}{\textbf{generalDeveloperForum}} & \multicolumn{1}{l|}{23813} & 500 \\ \hline
\multicolumn{1}{|l|}{\textbf{quizAndQuestionBanks}} & \multicolumn{1}{l|}{17269} & 500 \\ \hline
\multicolumn{1}{|l|}{\textbf{themes}} & \multicolumn{1}{l|}{10638} & 500 \\ \hline
\multicolumn{1}{|l|}{\textbf{generalPlugins}} & \multicolumn{1}{l|}{4866} & 500 \\ \hline
\multicolumn{3}{|l|}{\textbf{$\hdots$}} \\ \hline
\multicolumn{3}{|l|}{\textbf{other (52 forums | 76900 posts | 21213 Samples)}} \\ \hline
\multicolumn{3}{|l|}{\textbf{$\hdots$}} \\ \hline
\multicolumn{1}{|l|}{\textbf{poodLL}} & \multicolumn{1}{l|}{159} & 88 \\ \hline
\multicolumn{1}{|l|}{\textbf{journal}} & \multicolumn{1}{l|}{129} & 85 \\ \hline
\multicolumn{1}{|l|}{\textbf{artificialIntelligence}} & \multicolumn{1}{l|}{41} & 19 \\ \hline
\multicolumn{1}{|l|}{\textbf{moodleWorkplace}} & \multicolumn{1}{l|}{31} & 19 \\ \hline
\multicolumn{1}{|l|}{\textbf{userExperience}} & \multicolumn{1}{l|}{29} & 15 \\ \hline
\multicolumn{1}{|l|}{\textbf{Total}} & \multicolumn{1}{l|}{\textbf{204,289}} & \textbf{24,070} \\ \hline
\end{tabular}
\label{tab:communityForumDatasetNumbers}
\end{table}

To categorize the \textbf{feature requests}, we classify the sampled forum posts using GPT-4o~\cite{OpenAIGPTPage}
with the temperature parameter fixed at 0.0 to ensure deterministic and reproducible outputs. We select GPT-4o due to its strong instruction-following capabilities, robust in-context learning (ICL) performance, and efficiency in zero-shot text classification, and because it has been widely adopted in many prior works, making it a strong choice for this setting\cite{santos2024requirements, pragyan2025demystifying, bhatt2025automatic, korn2025llmrei}. The model is used in an ICL setting to perform classification using only the post title and description, thereby restricting the input to information available at the time the post was created (see Figure~\ref{fig:zero_shot_prompt_no_comments}). All prompts and responses are publicly available at the provided link~\cite{re2026_github_repo}. We adopt the following definition of a feature request: ``a post in which the author explicitly requests new functionality, missing functionality (e.g., features available in other systems), new content, or modifications and improvements to existing features in future releases. Posts that primarily report defects, seek troubleshooting assistance, ask usage or configuration questions, announce updates, or engage in general discussion are classified as non–feature requests''\cite{maalej2015bug}. 

We only process posts whose full content fits within the model context window; no truncation is applied, as truncation may remove information critical for classification. In our case, no inputs exceeded the context window. In total, we perform 24,070 API calls, with a total execution time of 7.94 hours. To this end, we have a corpus of community forum posts that are classified as feature requests. 

To evaluate classification accuracy, we conduct a human validation study using capped stratified sampling. For each forum, we randomly sample up to 25 posts; if fewer than 25 eligible posts are available (e.g., due to filtering for posts with images), we include the full set. This procedure yields a total of 1,528 unique posts across 62 forums, corresponding to 1,528 classification instances. The per-forum sample sizes used for validation are reported in~\cite{re2026_github_repo}. The first author manually validates all sampled instances according to the same feature request definition used in the prompts. Of the 1,528 validated instances, 1,446 are correctly classified, resulting in an overall accuracy of 94.63\%. 
The validation sample is used solely to assess classification accuracy. Misclassified instances identified during validation are not removed or corrected in the full dataset, as the validation set represents only a sampled subset of posts rather than the entire corpus.



\begin{figure}[t]
\centering
\fbox{
\begin{minipage}{0.95\linewidth}
\footnotesize
You are a requirements analyst auditing discussion forum posts for potential feature requests.

You will be provided with a Moodle discussion forum post that includes the post title and post description. Moodle is a free and open-source learning management system (LMS) used by schools, universities, and workplaces to create online courses and manage learning activities.

Your task is to classify if the provided discussion forum post is a feature request or not for Moodle. A feature request is a post where the user asks for new functionality, missing functionality (e.g., features available in other apps), new content, or changes/improvements to existing features in future releases. Posts that are not feature requests include: bug reports, troubleshooting questions, usage questions, configuration issues, announcements, or general discussions.

Respond strictly with one of the following labels: \texttt{FEATURE\_REQUEST} or \texttt{NOT\_FEATURE\_REQUEST}. If the forum post is a feature request, return \texttt{FEATURE\_REQUEST}. If the forum post is not a feature request, return \texttt{NOT\_FEATURE\_REQUEST}. Use the title of the post and the description of the post to make the assessment.
Respond strictly with one of the following labels:
\texttt{FEATURE\_REQUEST} or \texttt{NOT\_FEATURE\_REQUEST}.

Forum Post: \textbackslash n Post Title: \textbackslash n Post Description: \textbackslash n Label: 
\end{minipage}
}
\caption{Classification Prompt for Forum Posts}
\label{fig:zero_shot_prompt_no_comments}
\end{figure}

\subsubsection{Issue Tracker Feature Requests Dataset}\label{dataset:IssueTracker} 

We use Jira Query Language (JQL) queries with bounded date ranges and filters for \emph{feature request} and \emph{improvement} labels to export feature requests from Jira~\cite{AtlassianJIRA} in CSV format. We execute the queries over consecutive five-year intervals, starting from 2002, to ensure complete coverage of the project's issue history, yielding 23,169 feature requests. 

\subsubsection{Moodle Developers' Information Dataset}\label{dataset:DevInfo}
We collect developer-related information for Moodle from the publicly available \textit{Developer Credit}~\cite{MoodleContributorPage} page. For each developer, we extract the developer’s name, country, Moodle profile link, GitHub commit list, and GitHub merge list. In total, we collect information for 1,200 developers.

\subsection{Quantitative Method} \label{sec:quantitativeMethod}
We analyze the constructed datasets to measure and characterize cross-channel traceability links between forum posts and tracker issues, examine role-based authorship patterns in creating feature requests, and analyze the textual characteristics of feature requests to compare how they are articulated across channels within the Moodle project. 



\subsubsection{Traceability Analysis} \label{subsubsection:traceability}
Our goal is to identify and align issues with the community forum posts that motivated or informed them. To this end, we construct a corpus of \emph{issue-forum post pairs} that captures cross-channel traceability between community forum posts and issue tracker feature requests. We start from the \textbf{issue tracker dataset}, which contains 23,169 feature request issues. 
We identify traceability links by detecting explicit hyperlinks from issue descriptions to Moodle forum posts. Link detection is performed using pattern matching with the following regular expression: 
\texttt{https?://moodle.org/mod/forum/\allowbreak discuss.php?\allowbreak d=\textbackslash d+}. 
All other hyperlinks are excluded to ensure high precision in identifying trace links. After extraction, we post-process each candidate link to validate it by checking that (i) the URL resolves without HTTP 4xx/5xx errors and (ii) the referenced forum content is accessible (i.e., not removed or access-restricted). Only validated links are retained. Next, for each traced issue, we retrieve the corresponding forum post text from our \textbf{community forum dataset}. If an issue contains multiple traceable links, we select the forum post that is most semantically similar to the issue description using \texttt{all-MiniLM-L6-v2} from sentence embeddings~\cite{SentenceTransformer} and cosine similarity.
We choose \texttt{all-MiniLM-L6-v2} because it provides a strong trade-off between semantic accuracy and computational efficiency for large-scale text similarity comparisons~\cite{SentenceTransformer}.

The outcome of this process is a curated corpus of aligned \emph{issue-forum post pairs}, where each pair represents two linked texts describing the same feature request from the issue tracker and the community forum, respectively. 

\subsubsection{Authorship Analysis}
We analyze authorship roles across community forums and the issue tracker with the objective of understanding who initiates and maintains feature requests, and how different stakeholder groups contribute to requirements across these two channels. Specifically, we extract and analyze author role information from both the \textbf{community forum} \textbf{and} \textbf{issue tracker} \textbf{datasets}. 

In the \textbf{community forum dataset}, each forum post is associated with metadata that may include an official project role badge (e.g., \textit{core developer}, \textit{plugin developer}, \textit{tester}, \textit{translator}, among others). Across this dataset, we observe up to 11 distinct role labels. While these fine-grained labels reflect diverse forms of participation, several occur infrequently and are highly imbalanced. Analyzing them separately would therefore lead to sparse groups, unstable estimates, and reduced statistical power.

To address this, and to improve robustness and interpretability of our analyses, we aggregate these labels into three higher-level stakeholder categories based on their functional responsibilities in the project: \textit{developer}, \textit{maintainer}, and \textit{user}. Table~\ref{tab:roles_into_categories} summarizes the mapping from the original role labels to these categories. When no role badge is displayed, the forum does not provide explicit role information for that account. In such cases, we conservatively classify the author as a \textit{user}.

\begin{table}[]
\centering
\caption{11 Distinct Roles in Moodle Community Forum}
\begin{tabular}{|c|ccc|c|}
\hline
\textbf{developer} & \multicolumn{3}{c|}{\textbf{maintainer}} & \textbf{user} \\ \hline
\begin{tabular}[c]{@{}c@{}}core\\ developer\end{tabular} & \multicolumn{1}{c|}{\begin{tabular}[c]{@{}c@{}}moodle\\ workplace\end{tabular}} & \multicolumn{1}{c|}{\begin{tabular}[c]{@{}c@{}}peer\\ reviewers\end{tabular}} & \begin{tabular}[c]{@{}c@{}}moodle \\ cloud\end{tabular} & \begin{tabular}[c]{@{}c@{}}particularly\\ helpful moodlers\end{tabular} \\ \hline
\begin{tabular}[c]{@{}c@{}}plugin\\ developer\end{tabular} & \multicolumn{1}{c|}{\begin{tabular}[c]{@{}c@{}}plugin\\ guardian\end{tabular}} & \multicolumn{1}{c|}{\begin{tabular}[c]{@{}c@{}}document\\ writer\end{tabular}} & translator & no roles \\ \hline
 & \multicolumn{1}{c|}{moodle HQ} & \multicolumn{1}{c|}{tester} &  &  \\ \hline
\end{tabular}
\label{tab:roles_into_categories}
\end{table}



This abstraction preserves the key distinctions relevant to our research questions, namely, contributors who implement features, those who curate and manage the project, and end users who report needs, while reducing noise due to label sparsity and role fragmentation.

In the \textbf{issue tracker dataset}, Moodle’s Jira does not provide explicit role labels for issue reporters. We therefore leverage our \textbf{developer information dataset} to infer whether an issue was authored by a developer or a non-developer. Specifically, we extract developers' full names from the \textbf{developer information dataset} and match them against the reporter names in the \textbf{issue tracker dataset}\cite{hellman2022characterizing}. Based on this name-matching, we classify each issue as either \textit{developer-authored} or \textit{non-developer-authored}. 

We apply the same name-matching procedure to each \emph{issue–forum post pair} to determine whether the pair was authored by the same individual.

\subsubsection{Textual and Readability Analysis}
We examine how feature requests are articulated across community forums and the issue tracker by comparing the textual characteristics of \textit{issue–forum post pairs} in terms of length, linguistic readability, and domain-specific technical complexity. 
For each pair, we compute four complementary metrics: cosine similarity~\cite{morales2024large}, word count, Flesch-Kincaid Reading Ease (FKRE)\cite{kincaid1975derivation}, and a GPT-based technical complexity score (hereafter GPT Complexity Score)\cite{cachola2025evaluating}. Cosine similarity quantifies the semantic overlap between each forum post and its corresponding issue description. Word count captures differences in verbosity between forum posts and issue descriptions. FKRE provides a standardized, surface-level measure of linguistic readability. The GPT Complexity Score estimates the level of technical knowledge required to understand and act on each Moodle-related text, thereby capturing domain-specific technical complexity (e.g., references to Moodle internals, plugins, configuration, logs, or code) that is not fully reflected by traditional surface-level readability metrics. This choice is motivated by recent evidence that ICL–based models capture perceived technical difficulty more accurately than classical readability formulas\cite{cachola2025evaluating}. In our setting, we utilize GPT-4o with the temperature set to 0.0 for deterministic results and prompt it to act as a Moodle support expert (see Figure~\ref{fig:gpt_readability_score}) to assign an integer score from 1 (plain language understandable by non-technical users) to 5 (requiring advanced technical knowledge). 

We report average cosine similarity and word count, along with the mean, median, and standard deviation for FKRE and the GPT Complexity Score. Because the distributions for FKRE and GPT Complexity Score are non-normal and skewed, we use the Mann--Whitney U test\cite{mann1947test} to compare issues and forum posts. We report p-values to assess statistical significance and rank-biserial correlation\cite{kerby2014simple} as an effect size to quantify the magnitude and direction of the observed differences, enabling interpretation of both statistical and practical significance. Our null hypothesis is that forum posts and issue descriptions do not differ in readability or textual complexity.

\begin{figure}[t]
\centering
\fbox{
\begin{minipage}{0.95\linewidth}
\footnotesize
You are an expert Moodle support analyst. 
Moodle is a free and open-source learning management system (LMS) used by schools, universities, and workplaces to create online courses and manage learning activities.
Your task is to evaluate Moodle-related texts such as an issue from Moodle's issue tracker (Jira) or a post from Moodle's community discussion forum and assign a single score that represents how much technical knowledge of Moodle is required to understand and act on the text. This score, called technical\_knowledge\_required, must be an integer from 1 to 5, where 1 means the text can be understood by any non-technical Moodle user (student/teacher) using plain language, and 5 means the text requires advanced technical knowledge such as understanding Moodle internals, PHP, plugins, stack traces, database or server configuration, logs, or code. A score of 3 means moderate technical knowledge is needed (e.g., basic admin concepts or some technical terms, but still partly understandable to non-experts). Return ONLY valid JSON with the following fields: technical\_knowledge\_required (integer 1–5) and reasoning (an array of 2–4 short strings explaining the score, e.g., ``Contains stack trace", ``Uses plugin configuration terms", ``Plain user language").
Moodle related text: 
\end{minipage}
}
\caption{Prompt to find GPT Complexity Score}
\label{fig:gpt_readability_score}
\end{figure}



\subsection{Qualitative Method}

To complement the quantitative analysis, we conduct two semi-structured interviews with users and developers involved in the Moodle project. For each interview, we begin with demographic questions and then proceed to Interview Questions (IQs). We also ask follow-up questions as needed to properly address IQs. The interviews are reviewed and approved by the Institutional Review Board (IRB).

\noindent\textbf{Developer Interviews:}  
We conduct developer interviews with a goal to understand how feature requests are received, clarified, and refined by developers; how prioritization decisions are made; and how such requests are managed within the issue tracking system and community forums, including the creation and maintenance of trace links. The interview questions (D-IQ.X) for developer interviews are presented in Table \ref{tab:developerInterviewQuestions}.

\begin{table}[]
\caption{Developer Interview Questions (D-IQ)}
\begin{tabular}{|c|l|}
\hline
\textbf{Categories} & \multicolumn{1}{c|}{\textbf{D-IQs}} \\ \hline
\multirow{2}{*}{\textbf{\begin{tabular}[c]{@{}c@{}}Request \\ Intake\end{tabular}}} & \textbf{\begin{tabular}[c]{@{}l@{}}D-IQ.1.1 How do you usually find out about \\ feature or enhancement requests that come \\ from community discussion forums?\end{tabular}} \\ \cline{2-2} 
 & \textbf{\begin{tabular}[c]{@{}l@{}}D-IQ.1.2 What are the details included in \\ the received feature request?\end{tabular}} \\ \hline
\multirow{3}{*}{\textbf{\begin{tabular}[c]{@{}c@{}}Communication \\ and \\ Acknowledgement\end{tabular}}} & \textbf{\begin{tabular}[c]{@{}l@{}}D-IQ.2.1 When do you communicate with \\ users/requesters about the feature request?\end{tabular}} \\ \cline{2-2} 
 & \textbf{\begin{tabular}[c]{@{}l@{}}D-IQ.2.2 What do you ask users/requesters \\ about the feature request when \\ communicating?\end{tabular}} \\ \cline{2-2} 
 & \textbf{\begin{tabular}[c]{@{}l@{}}D-IQ.2.3 Where do you usually communicate \\ with the requester?\end{tabular}} \\ \hline
\textbf{Prioritization} & \textbf{\begin{tabular}[c]{@{}l@{}}D-IQ.3.1 What factors influence prioritization \\ e.g., technical complexity, alignment with project \\ goals, or community demand?\end{tabular}} \\ \hline
\multirow{3}{*}{\textbf{\begin{tabular}[c]{@{}c@{}}Traceability \\ and \\ Evolution\end{tabular}}} & \textbf{\begin{tabular}[c]{@{}l@{}}D-IQ.4.1 Do you use any tagging, linking, or \\ tracking mechanisms between community \\ forums and issue trackers?\end{tabular}} \\ \cline{2-2} 
 & \textbf{\begin{tabular}[c]{@{}l@{}}D-IQ.4.2 When a feature idea comes from a \\ community forum, how does it get turned \\ into an issue in your tracker?\end{tabular}} \\ \cline{2-2} 
 & \textbf{\begin{tabular}[c]{@{}l@{}}D-IQ.4.3 What practices or tools could \\ make it easier to connect user feedback \\ with development workflows?\end{tabular}} \\ \hline
\end{tabular}
\label{tab:developerInterviewQuestions}
\end{table}

Using the commit links and email addresses included in the \textbf{developer information dataset}, we invite developers who have made at least one commit within the past three years, yielding a pool of 112 eligible developers. To increase the response rate, we send the invitation emails twice, with a gap of 20 days between the two rounds. As a result, we conduct eight interviews via Zoom, where participants receive a \$50 gift card upon completion. 

\noindent\textbf{User Interviews:} 
We also conduct interviews with Moodle users, defined as individuals who participate in the Moodle community forum. We aim to understand how users communicate feature requests, how they perceive acknowledgment of their requests, and how they experience developer feedback and follow-up throughout the request process. The user interview questions (U-IQ.X) are listed in Table~\ref{tab:userInterviewQuestions}. 

\begin{table}[]
\caption{User Interview Questions (U-IQ)}
\begin{tabular}{|l|l|}
\hline
\textbf{Category} & \textbf{User Interview Questions (U-IQ)} \\ \hline
\textbf{\begin{tabular}[c]{@{}l@{}}Request Submission\\ and \\ Content\end{tabular}} & \textbf{\begin{tabular}[c]{@{}l@{}}U-IQ 1.1 Which channels do you use \\ to communicate feature requests \\ (e.g., discussion forums, issue trackers,\\ or elsewhere)?\\ U-IQ 1.2 How easy is it to navigate \\ through these channels?\\ U-IQ 1.3 What information do you \\ usually include in your feature request?\end{tabular}} \\ \hline
\textbf{\begin{tabular}[c]{@{}l@{}}Follow-up\\ and \\ Traceability\end{tabular}} & \textbf{\begin{tabular}[c]{@{}l@{}}U-IQ 2.1 After submitting a feature \\ request, how do you usually check \\ whether it has been acted upon?\\ U-IQ 2.2 Who transfers the user \\ requests to a developer (if anyone)?\end{tabular}} \\ \hline
\textbf{\begin{tabular}[c]{@{}l@{}}Communication\\ and Feedback\end{tabular}} & \textbf{\begin{tabular}[c]{@{}l@{}}U-IQ 3.1 Have you ever received any \\ clarification questions or feedback on \\ your request from developers, maintainers, \\ or other users?\\ U-IQ 3.2 Where do developers \\ communicate with you for clarification, \\ if any?\\ U-IQ 3.3 How responsive do you feel \\ developers or maintainers are to user \\ requests?\\ U-IQ 3.4 What would make it easier or\\ more effective for you to propose a feature \\ request?\end{tabular}} \\ \hline
\end{tabular}
\label{tab:userInterviewQuestions}
\end{table}


We utilize the \textbf{community forum dataset}, which includes user profiles associated with posts in its metadata. We obtain 225 email addresses of users who have posted on the community forum. To increase the response rate, we send the invitation email twice, with a gap of 25 days between the two rounds. We receive 12 responses, out of which 8 participants complete the interview via Zoom. To be eligible for participation, users are required to have submitted at least one feature request post on the community forum. 
Upon completion, participants receive a \$25 gift card. 
\section{Results} \label{sec:results}

\subsection{Quantitative Analysis}

\noindent\textit{(1) Community Forum Feature Requests-} 
From the 24,070 posts sampled from \textbf{community forum dataset}, we classify 5,163 posts as feature requests. The highest ratio of feature requests (relative to the total number of posts) is observed in futureMajorFeature (80.56\%), while the lowest ratio is observed in hardwareAndPerformance (2.80\%). 


\noindent\textit{(2) Traceability Analysis-} 
Out of 23,169 feature request issues from \textbf{issue tracker dataset}, 818 issues contain at least one valid traceable link to a community forum post in their descriptions. Among these, 707 issues include a single link, while 111 issues include multiple links; in the latter case, we select the most semantically similar \textit{issue–forum post pair} for analysis. 
Furthermore, among the 818 \textit{issue–forum post pairs}, we find that 436 issues refer back to community forum posts that are classified as feature requests.

\noindent\textit{(3) Authorship Analysis-} 
Using our \textbf{community forum dataset}, we analyze authorship across 204,289 posts. Of these, 4,295 posts are authored exclusively by developers, 2,237 by maintainers, 5,731 by individuals holding both developer and maintainer roles, and 192,026 by users. We further disaggregate authorship by specific roles: core developers (7,214), plugin developers (9,177), moodle HQ (157), moodle workplace team (80), moodle cloud team (21), plugin guardians (225), peer reviewers (2,290), testers (4,410), documentation writers (3,110), translators (2,863), particularly helpful moodlers (9,409), and users with no role assigned (189,980).

Using the \textbf{issue tracker dataset}, we analyze 23,169 feature request issues. Developers author 12,250 of these (52.8\%), while the remaining 10,919 are submitted by non-developers. Of the 818 \textit{issue–forum post pairs}, 316 are authored by developers and the remaining 502 by non-developers. Additionally, 230 feature request issues share the same author name between the issue and the corresponding community forum post, based on exact string matching of full names. Among these 230 authors, 110 are developers. 

\noindent\textit{(4) Textual and Readability Analysis-} 
The average cosine similarity between \textit{issue–forum post pairs} is 0.668, indicating substantial textual overlap between feature request issues and their corresponding community forum posts. The average word count for feature request issues is 152.64, while the average word count for community forum posts is 168.49. 
For feature request issues, the mean FKRE score is 40.93, the median is 43.85, and the standard deviation is 18.19. For community forum posts, the mean FKRE score is 58.95, the median is 60.15, and the standard deviation is 12.46. A Mann–Whitney U test indicates a statistically significant difference between the two distributions ($U = 126{,}707.0$, $p < .001$), with a large effect size as measured by the rank-biserial correlation ($r_{\mathrm{rb}} = 0.62$).
For feature request issues, the mean GPT Complexity Score is 3.24, the median is 3.0, and the standard deviation is 1.01. For community forum posts, the mean score is 2.77, the median is 3.0, and the standard deviation is 0.98. A Mann–Whitney U test again shows a statistically significant difference between the two distributions ($U = 419{,}309.5$, $p < .001$), with a small-to-moderate effect size ($r_{\mathrm{rb}} = -0.25$).




\subsection{Qualitative Analysis} 
\noindent\textbf{(1) Developer Interviews-} 
We analyze eight semi-structured interviews with Moodle developers using the coding scheme derived from D-IQs in Table~\ref{tab:developerInterviewQuestions}. The participants represent a diverse group in terms of background and experience: they are based across multiple countries (including India, the UK, Finland, Canada, Belgium, Switzerland, Germany, and France), with ages ranging from early 30s to over 60. Their educational backgrounds span from high school to bachelor’s and master’s degrees, and their development focus includes back-end, DevOps, and full-stack work. In terms of Moodle experience, participants report between 2 and 15+ years of usage, and contribute across core and plugin development.  The interviews last an average of 43 minutes. %

\noindent\textbf{Request Intake and Content (D-IQ~1.1, 1.2)} Developers most commonly report learning about feature requests through issue tracker monitoring (n=7) and direct stakeholder communication (n=6). In contrast, only one developer reports directly monitoring community forums, and one mentions self-initiated feature requests. 

Regarding the content of received feature requests, most developers describe receiving high-level intent descriptions rather than detailed specifications (n=7). Only one participant reports regularly encountering technical implementation details in incoming requests, and one reports limited or indirect visibility into user-originated feature requests. 

\noindent\textbf{Communication and Acknowledgment (D-IQ~2.1--2.3)} Developers report engaging in several forms of communication with users around feature requests, most commonly clarification-seeking (n=4), followed by reformulations (n=2) and intent elicitation (n=1). One participant reports no direct communication with users. 
When asked what information they typically request from users, three developers report requesting no additional information. Others describe asking for usage context (n=2), reproduction steps or supporting evidence (n=2), scenario or use-case descriptions (n=1), missing constraints or details (n=1), and supporting artifacts such as logs or screenshots (n=1). Communication with users primarily occurs through the issue tracker (n=5) or via direct communication channels (n=3), with one case involving internal, team-mediated communication. 

\noindent\textbf{Prioritization Practices (D-IQ~3.1)} Developers report self-driven prioritization (n=3), followed by effort-based (n=2) and organization-driven prioritization (n=2). Developers also mention technical complexity-based (n=1) and impact-based prioritization (n=1). 


\noindent\textbf{Traceability and Evolution (D-IQ~4.1--4.3)} 
Manual cross-platform search (n=3) and manual linking between platforms (n=3) are the most common practices, while three participants report no cross-platform linkage at all. One participant reports the use of internal-only traceability mechanisms. 
When describing how forum posts are turned into issue tracker entries, five developers report having no formal intake process. In contrast, three participants describe more structured approaches, including third-party mediated evolution (n=1), scenario-based reformulation (n=1), and structured requirements evolution (n=1). 
Finally, when asked about practices or tools that could improve the connection between user feedback and development workflows, responses were distributed across several themes: need for improved communication practices (n=3), perceived sufficiency of current processes (n=3), need for structured triage processes (n=1), clearer role ownership (n=1), and more accessible request platforms (n=1). 

\noindent\textbf{(2) User Interviews-} 
We analyze eight semi-structured interviews with Moodle users using the coding scheme derived from U-IQs in Table~\ref{tab:userInterviewQuestions}. The participants are based across multiple countries (including Canada, Switzerland, the USA, the Netherlands, and Japan), with ages ranging from early 40s to late 60s. Their educational backgrounds range from undergraduate and associate degrees to master’s and PhD degrees. In terms of roles, participants primarily reported using Moodle in teaching and administrative contexts. Their experience with Moodle spans from 10 to 22 years. The interviews last an average of 38 minutes. 

\noindent\textbf{Request Submission Channels and Navigation (U-IQ~1.1-1.3)} 
Three participants primarily rely on the community forum, two report using both the forum and the tracker, and three use the issue tracker as their main or only channel. Two participants also mention mediated or organizational channels (e.g., moderators or organizational processes) for submitting requests.
Perceived ease of navigation varies substantially across participants. Four participants report the channels as ``easy'' or ``navigable'', three report as ``overloaded'' or ``cluttered'', and one participant explicitly describes the system as ``not navigable.''
In terms of content, five participants report including intent and technical details in their requests, with four of these also adding links. The remaining responses show clear variation in detail: one participant includes scenario, intent, and links; one provides intent only; and one includes intent plus a forum link.

\noindent\textbf{Follow-up and Traceability (U-IQ~2.1-2.2)} Seven participants report relying on automated notifications to check whether their request has been acted upon. Only one participant mentions manual checking as part of their follow-up strategy. 
When asked who transfers user requests from the initial channel to a developer, responses are inconsistent. Two participants believe moderators perform this role, two indicate ``moderators or self,'' one reports ``self'' only, and three answer ``unknown'' or ``no one.'' 

\noindent\textbf{Communication, Feedback, and Perceived Responsiveness (U-IQ~3.1--3.4)} 
Four participants report receiving clarification questions, most often focused on intent or missing details. One participant reports receiving technical clarification, and one reports never checking for clarification. When clarification does occur, it takes place either in the issue tracker (n=3), the community forum (n=2), or both (n=1). Perceived responsiveness of developers and maintainers ranges from ``very responsive'' (n=1) and ``moderately responsive'' (n=2) to ``rarely responsive'' (n=3) and ``not responsive'' (n=2).
When asked what would make it easier or effective to propose feature requests, participants point to several needs: better navigation and discoverability, clearer distinction between forum and tracker, better onboarding for non-technical users, more direct submission paths, improved communication and acknowledgment, and a more centralized entry point for requests. 
\section{Discussion} \label{sec:disscussion}
We now answer the RQs based on the results. 


\noindent\textbf{RQ.1: Prevalence of Feature Requests in Moodle Forums-} 
Our results indicate that feature requests constitute a substantial share of Moodle forum activity. Of the 24,070 sampled forum posts, we find 5,163 classified as feature requests, meaning that approximately one in every five posts in the dataset concerns a request for new or improved functionality. This finding suggests that, while the forums primarily serve purposes such as support, troubleshooting, and general discussion, they also operate as an important channel through which users propose new features and enhancements that might otherwise go unnoticed by developers. 

\noindent\textbf{RQ.2: Cross-Channel Initiation and Evolution of Feature Requests-} The authorship analysis shows stakeholder engagement in both the community forum and the issue tracker. For the community forum, a vast majority of feature requests are authored by users rather than developers or maintainers. 
This reinforces the view of community forums as a user-facing entry point for non-technical users who are not familiar with issue trackers, where users articulate needs and desired changes in their own terms.

For the issue tracker, we observe that developers author more issues overall compared to non-developers (users and maintainers), accounting for 52.8\% of all feature request issues. This contrasts with the forum, where feature requests are predominantly authored by users. This difference in authorship patterns indicates that the issue tracker is more actively used by technically involved stakeholders who participate directly in development workflows. 

\noindent\textbf{RQ.3: Cross-Channel Traceability Between Forums and Jira-} The traceability analysis reveals that only \textit{a small fraction} of issue tracker entries explicitly link back to forum posts. Out of 23,169 issues in the Moodle Jira tracker, only 818 contain at least one valid traceable link to a forum discussion. 
This indicates that explicit cross-channel linking between the issue tracker and forums exists, but is relatively rare compared to the overall volume of issues. 
Among these linked cases, we observe that both developers and non-developers create issues referencing forum posts, and in 230 cases, the same individual appears to be the author in both channels. This suggests that some users and developers actively bridge the two channels themselves, carrying a post from the forum into the issue tracker. However, given the small proportion of linked issues overall, such practices do not appear to be systematic or pervasive. 
The results suggest that \textit{while traceability links between forums and the issue tracker do exist and are used in practice, they are the exception rather than the norm.} Most posts in forums are not explicitly connected to issues in the tracker, which implies that much of the context and rationale expressed in forums may remain disconnected from the official record of requirements and development decisions. 
The limited prevalence of explicit cross-channel links is important because traceability serves purposes beyond simple documentation. Cross-channel traceability preserves the rationale, stakeholder context, and negotiation history behind feature requests as they evolve from user-facing discussions into development artifacts. Without such links, developers may lose access to the original motivations and contextual information that prompted a request, while users lose visibility into how their feedback influenced development decisions. Traceability therefore supports transparency, accountability, and sustained stakeholder engagement throughout the requirements lifecycle, rather than functioning merely as a mechanism for connecting artifacts.
These findings contrast with prior work by Tizard et al.~\cite{tizard2022software}, who reported a significant number of forum posts in VLC and Firefox linking to issue trackers and product documentation. Unlike those consumer-oriented OSS ecosystems, Moodle operates within a multi-stakeholder educational environment that includes teachers, administrators, instructional designers, and developers with diverse technical responsibilities. This structural and role-based heterogeneity may shape how feature requests migrate across communication channels. Further investigation is needed to systematically examine differences in participant composition and communication norms across these ecosystems.


\noindent\textbf{RQ.4: Textual Differences Between Forums and Issues-} 
On average, forum posts are easier to read than issue descriptions, as indicated by higher FKRE scores and lower GPT Complexity Score. Forum posts are also slightly longer on average, although the difference in word count is relatively small. 
Together, these measures show that forum posts require less technical background to understand, whereas issues more frequently assume technical knowledge and use more specialized terminology. The moderate cosine similarity between paired forum posts and issue descriptions further indicates that, even when a forum discussion is linked to an issue, the content is not simply copied verbatim. Instead, information appears to be reformulated, condensed, or reframed when moving from the forum context to the issue tracker.

\noindent\textbf{RQ.5: Developer Practices in Interpreting and Prioritizing Feature Requests-} 
The results show that developers primarily receive feature requests through issue trackers and direct stakeholder channels rather than by directly monitoring community forums. This indicates that community input is typically filtered or transferred into development workflows before reaching developers, which is consistent with our quantitative findings showing weak cross-channel traceability. Participants described this intake process in concrete, tool-centered terms, for example noting that they rely on \textit{``feature request through Moodle tracker for our plugin''} and that they work \textit{``from issue tracker -- get details with issue (priority, assigned, and more). I go through the description and all,''} while another emphasized that \textit{``most of the things come from GitHub since it is plugin.''} Together, these accounts reinforce the central role of issue trackers as the primary gateway through which requests become visible and actionable for developers, rather than forums serving as a direct input channel.

Most requests are received as high-level descriptions rather than detailed specifications, which requires developers to interpret and refine the intent of the user. This is reflected in the prevalence of clarification-seeking and reformulation-driven communication observed in the interviews. Developers characterized this interpretive work as an active sense-making process: one explained, \textit{``I communicate in GitHub to get scenario to understand the feature request,''} while another described a more structured approach, stating, \textit{``for feature request, someone comes with an idea, walk back and see if I understand, can I generalize and make it open for other workflow, estimate effort, and mockups, feature would need all these.''} These statements illustrate that developers do not treat requests as fixed specifications, but instead translate user-facing narratives into more general, technically viable, and project-aligned representations.

Communication around requests mainly takes place in issue trackers or through direct messages, reinforcing the central role of these tools in negotiating scope and meaning. This also extends to feedback and closure, as one participant noted: \textit{``For tracker -- we notify people when it is complete and notify people about their suggestion.''} This suggests that the issue tracker functions not only as an intake mechanism but also as the primary medium signaling progress.

Prioritization is most often driven by developers’ own judgments, along with effort and organizational constraints, while impact- and complexity-based considerations are less prominent. Several participants described prioritization as strongly developer- or team-controlled, for example stating that \textit{``Prioritization completely up to developer,''} or that \textit{``The team that collects requests prioritize it and developer get what to work.''} Others emphasized pragmatic constraints, noting that \textit{``first priority is community attraction want it and we would build, second is size of the project -- too long then we will postpone.''} 

\noindent\textbf{RQ.6: User Submission and Feedback Experience in Feature Requests-} 
From the user perspective, feature request submission and follow-up are shaped by fragmented channels, uneven process transparency, and a heavy reliance on tool-mediated feedback. Users report using both community forums and issue trackers, yet their accounts suggest that the boundary between these spaces—and the pathway by which a request moves from an idea to a development artifact—remains unclear. Descriptions of navigation as ``\textit{cluttered},'' ``\textit{overloaded},'' or even ``\textit{not navigable}'' point to discoverability and onboarding as key barriers, particularly for less technical participants. One participant explicitly highlighted this disparity in familiarity with the system: ``\textit{for me, I have a clear idea of tracker and forum, but not for everyone, so there might be a better onboarding}.'' While some users can navigate the ecosystem effectively, others struggle to understand where and how to submit requests.

Users also vary widely in how much technical detail they provide. Some include concrete technical information and supporting links, whereas others submit only high-level descriptions. As one participant states ``\textit{descriptive text, no technical, no version}.'' This reinforces the role of the forum as a more accessible, user-oriented space, in contrast to the issue tracker, which implicitly expects more structured and technical input. This qualitative observation aligns with our quantitative results, which show that forum posts are easier to read and less technically complex than issue descriptions.

A particularly salient finding is users’ heavy reliance on automated notifications for follow-up. Rather than actively tracking requests across channels, users tend to depend on the platform to surface updates. However, because traceability links are infrequent and responsibilities for transferring requests are unclear, this reliance likely contributes to the perception of low responsiveness reported by several participants. This dissatisfaction is evident in reflections on past experiences with the tracker. One user explains, ``\textit{the reason why I stopped using the tracker---I have come to learn that issues exist, but I never get help; once in a while help from the forum, but from the tracker I can't recall any helpful response}.'' Another similarly contrasted bug handling and feature requests: ``\textit{they are responsive for bugs; for feature requests they are not much}.'' In the absence of notifications or visible responses, users may reasonably conclude that their requests have been ignored or deprioritized, even when they have been discussed or reformulated elsewhere.

The ambiguity surrounding who is responsible for moving requests from forums to the issue tracker—whether moderators, developers, or users themselves—highlights a socio-technical gap in the current process. This uncertainty surfaces in statements such as ``\textit{without deliberately posting, I don't know},'' and ``\textit{I don't know how it works---the structure is unclear}.'' From the user perspective, there is no clear ownership of this handoff, increasing the risk that requests fall through the cracks or lose their original context. This directly echoes our developer-side findings, where many developers reported the absence of a formal intake process and reliance on largely manual or ad hoc linking practices.

Finally, users’ suggestions for improvement—such as a centralized entry point, clearer forum–tracker distinctions, better onboarding, and more explicit acknowledgment—point to the need for both improved tool support and clearer processes. Rather than expecting users to understand a project’s internal workflows, the ecosystem could better support \emph{progressive evolution}: allowing users to begin with informal, intent-focused requests and then guiding or assisting their transition into trackable development artifacts.

\noindent\textbf{Impact on the Requirements Engineering (RE) Community-} 
Our findings have several implications for the RE community, particularly as RE is a collaborative process involving multiple stakeholders who interact to elicit, refine, prioritize, and implement requirements~\cite{montgomery2025mining}. First, the observed rarity of explicit traceability links between community forums and issue trackers highlights a persistent gap between informal, user-facing requirements discussions and development artifacts. Second, the systematic differences we observe in readability and technical complexity between forum posts and issue descriptions provide empirical evidence of a form of progressive refinement in practice. Third, the mixed experiences reported by users regarding feedback, acknowledgment, and follow-up underscore the importance of traceability not only for developers, but also for stakeholder communication and trust. 

This reinforces a long-standing challenge in RE: given the inherent ambiguity and incompleteness of feature requests~\cite{pragyan2025demystifying}, how can we systematically preserve requesters' intent as requests transition from community discussions into development workflows, and how can developers effectively engage requesters in iterative clarification and feedback? Our results suggest that this gap is not purely technical but socio-organizational, shaped by informal practices, unclear responsibility boundaries, and fragmented tool ecosystems. For the RE community, this highlights the need for structured, role-aware methods and tool support that facilitate the progressive refinement of user-authored requests into actionable development artifacts while maintaining traceability to their original context. Such mechanisms can strengthen transparency, accountability, and sustained stakeholder collaboration, positioning cross-channel traceability not merely as a linking mechanism, but as a foundational element of collaborative change management in OSS ecosystems.

\noindent\textbf{Cross-Channel Traceability as a Socio-Technical Coordination Problem-} 
Our findings suggest that the observed traceability gap is not solely a tooling issue, but a broader socio-technical coordination challenge. While only 818 of 23,169 feature request issues ($\approx$3.5\%) explicitly link back to community forum discussions, the interviews reveal complementary explanations for this limited traceability. Developers rarely monitor forums directly and instead rely on issue trackers or intermediary stakeholders, whereas users often assume that moderators, maintainers, or other community members are responsible for transferring requests into development workflows.
Taken together, these findings point to unclear ownership of cross-channel traceability and fragmented communication practices. As feature requests move from community discussions to issue trackers, they are often reformulated without explicit links to the original conversation. Consequently, developers may lose access to the rationale and context behind a request, while users may have limited visibility into how their feedback influences development decisions. These results suggest that improving cross-channel traceability requires not only better tool support, but also clearer processes and responsibilities for connecting user discussions with development artifacts.

\section{Threats To Validity} \label{sec:threats} 
\noindent\textbf{Internal validity} is the extent to which measured variables
cause observable effects in the data~\cite{yin2009case}. Our quantitative analyses rely on automated data collection, classification, and link extraction. Errors in web scraping, misclassification of forum posts as feature requests, or missed or incorrectly matched trace links could affect the results. To mitigate this, we used deterministic model settings, validated extracted links, and conducted a human validation of a stratified sample of classifications. For the qualitative component, interview responses may be influenced by recall bias or social desirability bias~\cite{althubaiti2016information}. We reduced this risk by using semi-structured interviews, asking concrete, experience-based questions, and triangulating interview findings with artifact analysis. The reliability of the analysis is further supported by deterministic model configurations, documented procedures, and the public availability of datasets, prompts, and replication materials.

\noindent\textbf{Construct validity} is the correctness of operational measures
used to collect data, build theory, and report findings from the data~\cite{yin2009case}, and the extent to which an observed measurement
fits a theoretical construct~\cite{anderson2005experimental}. 
Readability and technical complexity are approximated using FKRE and a GPT-based scoring approach, which, while complementary, remain proxies for human judgment. In addition, our estimates of the prevalence of feature requests in community forums are based on a capped, stratified sample rather than the full population of posts. While this strategy is intended to balance representation across forums, it may over- or under-estimate the true proportion of feature requests in the complete dataset. Furthermore, our analysis focuses primarily on Moodle's Jira issue tracker, while parts of the ecosystem (e.g., plugins) rely on GitHub and may follow different workflows, which means our constructs may not fully capture all requirements-related practices across the broader Moodle ecosystem. 

\noindent\textbf{External validity} determines the scope of environmental phenomena or domain boundaries to which the theory and findings generalize~\cite{yin2009case}. This study focuses on a single, large learning management OSS project. While Moodle is a rich and representative case of a mature, community-driven ecosystem, its processes, tools, and community norms may differ from those of smaller projects, OSS in other domains (e.g., VLC \& Firefox), or projects using other communication channels, such as GitHub issue tracker. As a result, our findings may not generalize directly to all OSS or industrial contexts.
In addition, recruiting interview participants from Moodle HQ and core maintainers proved challenging, which means their perspectives may be underrepresented in our qualitative results. Furthermore, our user interview participants are recruited from the Moodle community forum dataset and therefore represent individuals already engaged with forum-based communication. Users who primarily interact through issue trackers, organizational channels, or other communication mechanisms may have different experiences and perceptions. Consequently, the interview findings may overrepresent the perspectives of forum-active participants. 
This may limit the generalizability of our interview-based findings to the full range of stakeholder roles involved in Moodle's development and governance. Replications across different projects and ecosystems are needed to assess the broader applicability of our results.

\noindent\textbf{Conclusion validity} concerns the extent to which the statistical analyses support the conclusions drawn from the data. Our quantitative findings may be influenced by sampling decisions, role-classification heuristics, and automated measures of technical complexity. In addition, the forum post classification achieved an accuracy of 94.63\%, indicating that some feature requests may have been misclassified. To mitigate these risks, we report effect sizes alongside significance tests, employ non-parametric statistical methods appropriate for non-normal distributions, and triangulate quantitative findings with qualitative interview data. The consistency between the quantitative and qualitative results increases confidence that the observed patterns are not artifacts of a single data source or analytical method.


\section{Conclusion} \label{sec:conclusion}
This paper presents a mixed-methods case study of cross-channel traceability between community forums and issue trackers in the Moodle open-source ecosystem. By combining large-scale artifact analysis with interviews of developers and users, we examined how feature requests are authored, transformed, and linked across two channels. 
Our results show that explicit traceability between forums and the issue tracker is rare, that users and developers occupy different roles in initiating and maintaining feature requests, and that forum posts and issue descriptions differ systematically in readability and technical complexity. The qualitative findings further reveal that the transition from forum posts to issues is largely ad hoc, with limited tool support and unclear role ownership, and that users often experience the process as opaque or weakly responsive. 
Taken together, these findings highlight a persistent gap between user-facing requirements discussions and development-centric artifacts, underscoring the need for better support in traceability and feedback loops. 
In future work, we will replicate our analyses in diverse OSS projects,  examining differences in participant composition, governance structures, and communication norms to understand how structural and role-based heterogeneity shapes cross-channel traceability practices. Additionally, we will explore tool- and process-level interventions that more tightly integrate community discussions with requirements management workflows. 

\section{Acknowledgment}
The authors thank Dr. Travis Breaux for his suggestions and insights. This work is supported by NSF award \#2318915.

\section{Data Availability Statement}
The dataset used in this study is publicly available at ~\cite{re2026_github_repo}. To protect the privacy and anonymity of individuals associated with the forum posts, usernames, email addresses, and other personally identifiable information were removed or redacted prior to publication. The resulting de-identified dataset is available to support the reproducibility of the reported findings and facilitate future research.

\bibliographystyle{IEEEtran}
\bibliography{reference}

\end{document}